\documentclass{eas}
\usepackage{graphicx}
\usepackage{epsfig}

\begin{document}

\title{From stars to nuclei} 
\author{Georges Meynet}\address{Geneva Observatory, CH--1290 Sauverny, Switzerland}
\begin{abstract}
We recall the basic physical principles governing the evolution of stars with some emphasis on the role played by the nuclear reactions. We argue that in general it is not possible from observations of stars to deduce constraints on the nuclear reaction rates. This is the reason why precise measurements of nuclear reaction rates are a necessity in order to make progresses in stellar physics, nucleosynthesis and chemical evolution of galaxies. There are however some stars which provides useful constraint on nuclear processes.  
The Wolf-Rayet stars of the WN type present at their surface
CNO equilibrium patterns. There is also the particular case of the abundance of $^{22}$Ne
at the surface of WC stars.
The abundance of this element is a measure of the initial CNO content. Very interestingly,
recent determinations of its abundance at the surface of WC stars tend to
confirm that massive stars in the solar neighborhood have initial metallicities in agreement with the
Asplund et al. (2005) solar abundances
\end{abstract}
\maketitle


\section{Stellar evolution in a nutshell}

\subsection{Luminosity as a consequence of hydrostatic equilibrium}

Stars are a privileged place in the universe where microscopic phenomena
interact with macroscopic ones. If the long range force of gravity plays the main
role in driving the birth of the stars, their life and sometimes their death (in core collapse supernovae), the other three interactions of physics contribute to the processes of production, transfer and loss of energy either under the form of electromagnetic radiation or through neutrinos.

During the longest part of their life, stars are in hydrostatic equilibrium.
An element of star in equilibrium undergoes a gravitational force balanced by
a pressure gradient.
In a normal star, pressure depends on temperature and therefore the existence
of a pressure gradient implies the build up of a temperature gradient.
Higher is the temperature, greater is the quantity of energy contained per 
unit volume in the radiation field.
Since the central parts are hotter than the outer ones due to the temperature gradient implied by hydrostatic equilibrium, energy flows from the inner parts of the star to the outer ones and thus the star continually loses energy. Interestingly, more efficiently the energy is evacuated from the central regions where it is produced, hotter become these central regions! Indeed, if the central regions loses energy, the temperature gradient will become weaker and the central regions will slowly contract, making them warmer!
Stars (made of perfect gas) are thus systems with a global negative specific heat.

This energy has of course to be extracted in one way or another from the stellar material.
These mechanisms of extraction of the energy are responsible for the evolution 
of the star.
Stars evolve because they continually lose energy to maintain hydrostatic equilibrium, or in other words to balance
the gravitational force.

It is interesting to note here that these logical deductions do not involve
any particular source of energy. Luminosity is a direct consequence of the
hydrostatic equilibrium and
not of the nuclear reactions which occur in its stellar interior.
Of course nuclear reactions are important and we are going to see how below.

\subsection{The main energy reservoirs}

There are actually two main sources of energy in a star: First,
a star can extract energy from the gravitational potential through global (macroscopic) contraction.
Second, a star can also 
release energy by the thermonuclear reactions which take place in its central regions where the temperature and density are adequate for such processes to occur (microscopic contractions).

These two processes for producing energy have different characteristic time\-scales.
If the Sun had only the gravitational energy source, its
lifetime would be of the order of a few tens of million years, instead a simple
estimate gives a lifetime of ten billion years when nuclear sources are 
taken into account. Only this last value is in agreement with what we know about the past history of Earth
and on the apparition of life at its surface.
Therefore the presence of the nuclear source is important not to explain the luminosity of stars
(stars could be shining even without hosting nuclear active regions), but for explaining why they can shine
{\it for very long durations}. Beside this energetic aspect, nuclear reactions are of course the process through which chemical elements are transformed in stars. They are thus at the heart of stellar nucleosynthesis and play a key role in the long chain of processes going from the Big Bang to the apparition of living bodies.
 
The evolution of a star can be viewed as a succession of phases where the energy is mainly produced by the nuclear reactions and of phases where
the energy is mainly produced by contraction.
When the star has burned all the nuclear fuel available in its central regions,
in order to maintain the luminosity required by the hydrostatic equilibrium, it has
to produce energy via contraction, until new central conditions are reached
adequate for the ignition of new nuclear reactions.
This succession of contraction periods will increase both the central temperature and density. At a given
point (which depends mainly on the initial mass of the star), the central regions can become sensitive to
degeneracy effects. 

\subsection{Energy production in perfect gas and degenerate conditions}

Let us recall that degeneracy pressure results from the exclusion principle:
only two fermions of spin one half, such as electrons, neutrons, or neutrinos,
may locally occupy the same quantum state.
Two particles are in two different quantum states if the product of their difference in position $\Delta x$ and their difference in momentum $\Delta p$ is
superior to the Planck constant.
An increase of the density restricts the domain for the positions and thus reduces $\Delta x$.
The exclusion principle means therefore that certain particles will acquire very
large impulses, much greater than that they would acquire by thermal agitation.
These particles with large velocities exert a new sort of pressure which is not
thermal in origin and which depends only on the density.

Depending on the fact that the stellar material
is degenerate or not, the two main sources of energy, contraction and
nuclear reactions, have very different behaviors. In non degenerate conditions, nuclear reactions are stable
and contraction implies an increase of the central temperature. In degenerate conditions, nuclear reactions
are explosive and contraction may produce a cooling of the medium.

Let us first consider a uniform contraction of a mass $M$. In that case a variation in radius $\Delta R$
corresponds to a variation in pressure $\Delta P$ and to
a variation in density $\Delta \rho$ so that we have the following relations:
$${\Delta P \over P}= -4 {\Delta R \over R}, \ \ {\rm and}\ \  {\Delta \rho \over \rho}=-3 {\Delta R \over R}.$$
The first equality is deduced from the hydrostatic equilibrium equation and the second from the continuity equation.
From these two relations, we can write
$$\Delta\ln P= {4 \over 3} \Delta\ln \rho.$$
Let us now write the equation of state as follows
$$\Delta \ln \rho=\alpha \Delta \ln P - \delta \Delta \ln T,$$
where $\alpha$ and $\delta$ are defined by $\alpha=\left({\partial \ln \rho \over \partial \ln P}\right)_{T,\mu}$ and
$\delta=-\left({\partial \ln \rho \over \partial \ln T}\right)_{P,\mu}$, and where $\mu$,
the mean molecular weight, is supposed to remain constant.
From these two relations one obtains, by eliminating $\Delta P$ the two following
relations between a variation in log T and log $\rho$:
$$\Delta\ln T=\left({4\alpha-3 \over 3\delta} \right)\Delta\ln \rho.\eqno(1)$$
For a perfect gas law we have $\alpha=\delta=1$. Therefore an increase of, for instance, 30\% in density implies an increases of 10\% in temperature. 

In the case the gas may be considered as completely degenerate and non relativistic we
have the following proportionality between the degenerate electronic gas pressure and density $P \propto \rho^{5/3}$.
In the log T versus log $\rho$ diagram, the region where the perfect gas pressure
for the electrons equal that for the degenerate gas is a line whose slope is equal to two third.
As seen above, the track of a slowly contracting
star in the central log temperature-density diagram is a straight line with slope one third
(see Eq.~(1) with $\alpha=\delta=1$).
Thus eventually a star will cross the frontier between the perfect and degenerate gas.
For a degenerate , non-relativistic gas, one has $\alpha=3/5$ and $\delta =0$.
Of course in this case, Eq.~(1) above is no longer valid, but if
during the course of evolution, when the central conditions pass from the non-degenerate region to the degenerate one, $\alpha$ becomes inferior to three quarters before $\delta$ is
equal to zero, then a contraction can produce a cooling! This can be understood as due to the fact that,
in order to allow electrons to occupy still higher energy state, some energy has to be extracted from the
non degenerate nuclei which, as a consequence, cool down.

Let us now study the nuclear source in degenerate conditions.
Let us imagine that for whatever reason an excess of energy is produced at the center of the star.
This will produce a heating of the matter. When the perfect gas law prevails, an
increase of temperature will produce an increase of pressure and therefore an expansion.
This implies an increase of the potential energy and through the Virial
theorem a decrease of the internal energy, therefore the temperature decreases as well as the nuclear reaction rates. We see that in perfect gas conditions,
there is a negative feedback which stops the runaway.
The nuclear reactions are stable when the perfect gas law prevails.

When the matter is degenerate, the behavior is quite different.
The excess of energy produced at the center, which implies an increase of the temperature
does no long provoke an expansion, since there is no long a coupling between 
pressure and temperature.
The nuclear reaction rates increase, new excesses of energy are produced, a 
flash or an explosion occurs.
The nuclear reactions are unstable in degenerate matter.
This process is responsible for the explosion of type Ia supernovae. It triggers also
what is called the helium flash at the tip of the Red Giant Branch for stars with masses below about
1.8 M$_\odot$ at solar metallicity.
These different behaviors of both contraction and nuclear reaction rates
in perfect gas and degenerate regimes are the main causes for the different evolution
followed by stars of different initial mass.
Schematically four ranges of masses are considered:
\vskip 1mm
\noindent 1) {\it Substellar objects}: This range includes objects that during their contraction phase enter the degenerate regime
before the ignition of hydrogen. Further contraction cools the central regions as explained above and
thus these central regions never reach the appropriate conditions for the ignition of hydrogen.
These objects become brown dwarfs. According to Baraffe et al. (2002) the stellar/substellar transition occurs for a mass around 0.075 M$_\odot$. Note that in objects more massive than 0.012 M$_\odot$
(Saumon et al. 1996; Chabrier 2000) deuterium burning occurs.
\vskip 1mm
\noindent 2) {\it The low mass stars}:This range includes stars that go only through the hydrogen burning
phase (progenitors of helium white dwarfs). These stars have initial masses between 0.075 and 0.5 M$_\odot$.
Note that single stars in the above mass range have Main-Sequence lifetimes greater
than the age of the Universe. Thus the observed helium white dwarfs likely originate from more
massive progenitors in close binary systems (see e.g. Benvenuto \& De Vito 2004).
\vskip 1mm
\noindent 3) {\it The intermediate mass stars}:This range includes stars that go only through the hydrogen and
helium burning phases (progenitors of carbon-oxygen white dwarfs). They have
initial masses between 0.5 and about 8 M$_\odot$. Observing white dwarfs in stellar clusters
allow to deduce the initial masses of the stars having given birth to the white dwarfs. Relations between
the mass of the white dwarf and the initial mass of the progenitors can be deduced. Such observations bring
very interesting constraints on the evolution of stars in this mass range, especially on the mass loss
undergone by these stars along the Red and Asymptotic Giant Branches (see Kalirai et al. 2007).
\vskip 1mm
\noindent 4) {\it The massive stars}:This range consists of all stars evolving beyond
the helium burning phase (progenitors of neutron stars and black holes). 
Generally massive stars have initial masses greater than about 8 M$_\odot$ 
(see e.g. Maeder \& Meynet 1989).

\section{The nuclear reactions rates in stellar models}

The various inputs of stellar models can be classified in two main categories:
the ingredients which comes from laboratory experiments and theoretical
considerations and quantities which comes from astronomical observations.
In the first class, one finds the opacities, the equation of state, the
nuclear reaction rates, the neutrino emission rates. In the second category
there are the initial mass of stars, their initial composition, their initial
rotation velocity, the mass loss rates. 
The convection parameter are in some respect in between these two classes of parameters. 
Ideally one would like to have an ab initio theory of convection enabling us 
to predict the turbulence from whatever physical conditions. Since
this theory does not exist, most stellar modelers use a phenomenological approach, 
introducing some free parameters calibrated on observed quantities sensitive to them.
Here we just comment the case of the nuclear reaction rates.

\begin{figure}
\includegraphics[width=2.5in,height=3in]{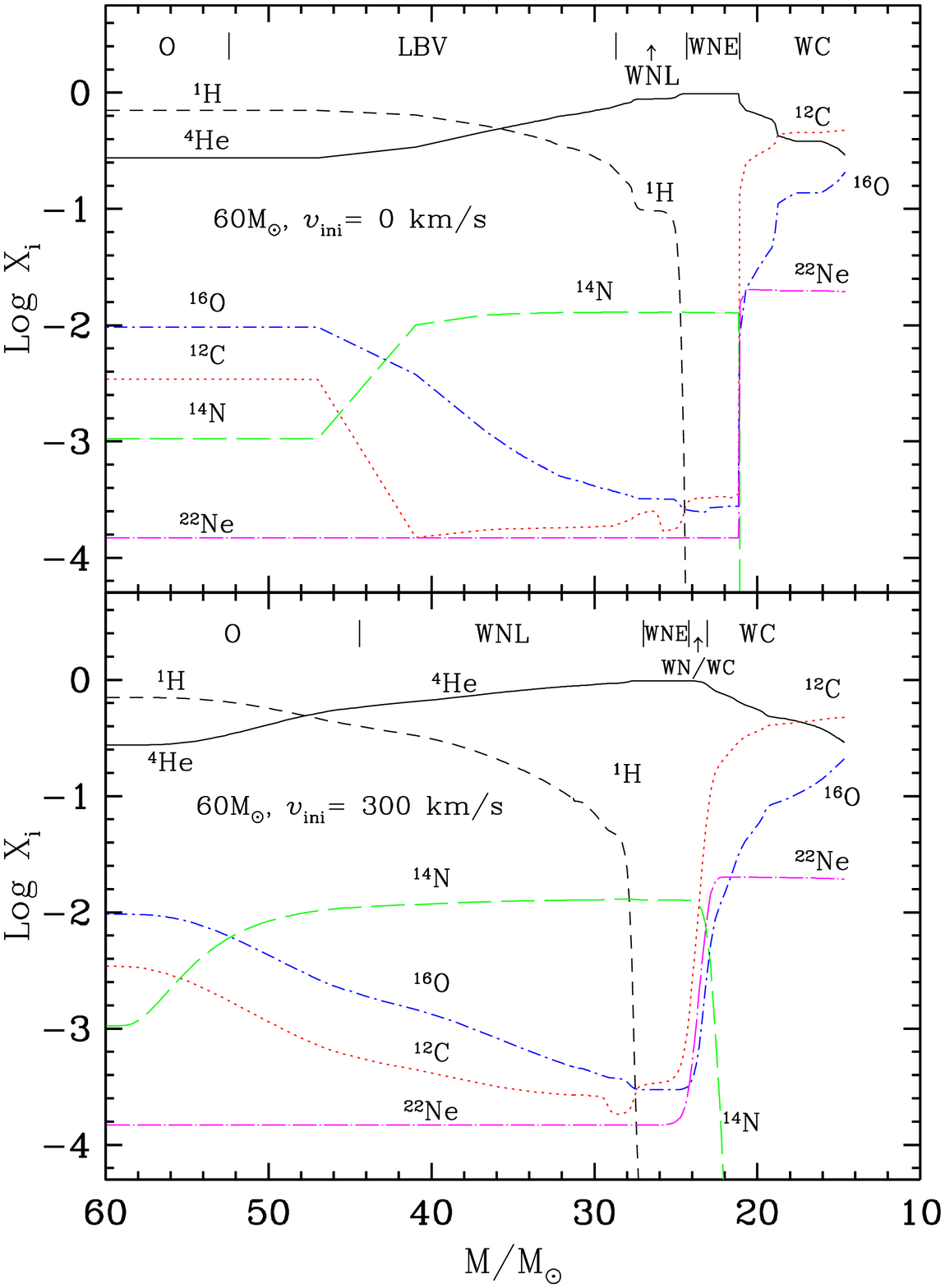}
\hfill
\includegraphics[width=2.5in,height=2.5in]{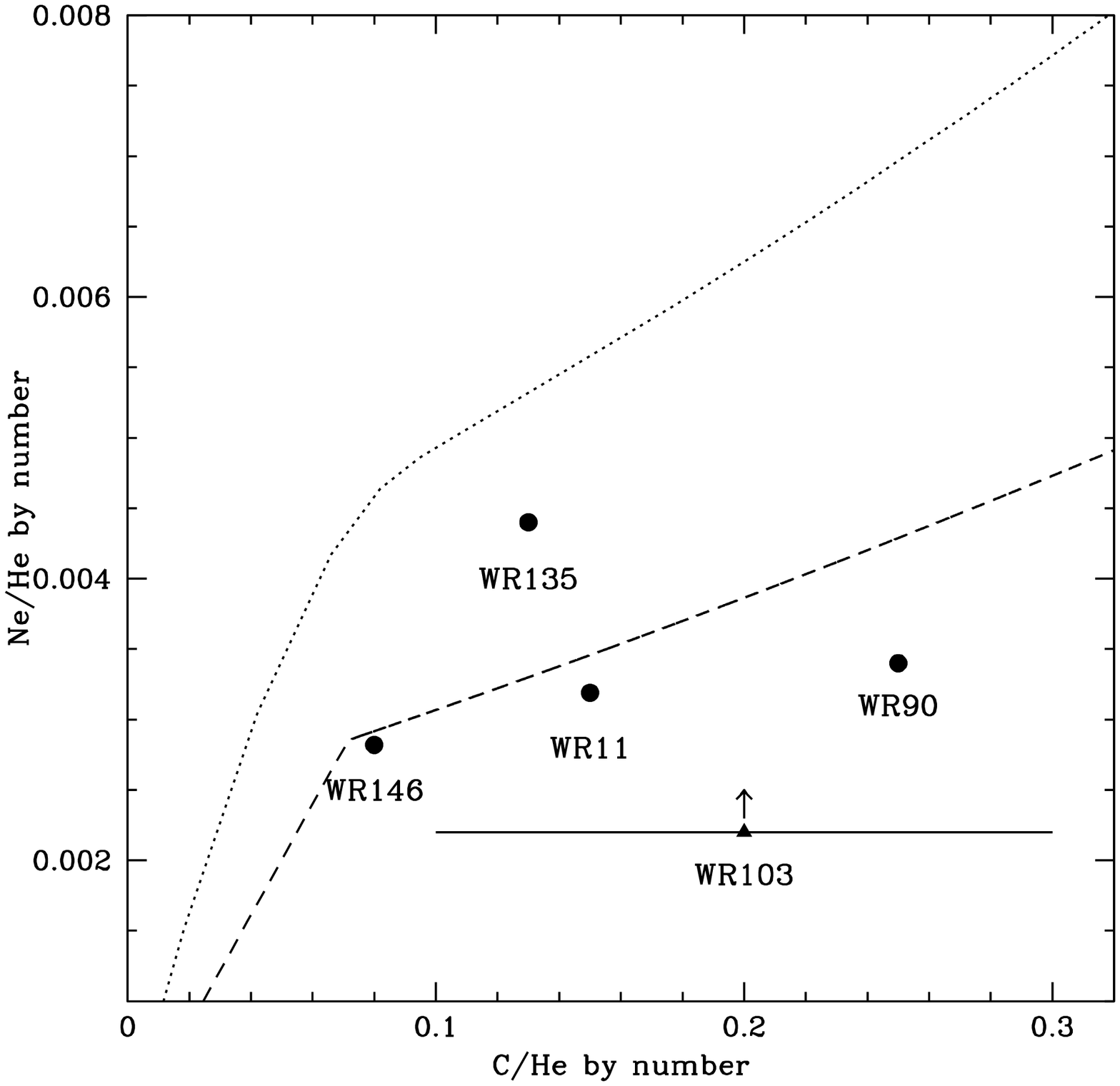}
\caption{{\it Left panel}:Evolution as a function of the actual mass of the abundances (in mass fraction) at the surface of a non--rotating  
(upper panel) and a rotating (lower panel) 60 M$_\odot$ stellar model. The actual mass of the star decreases as a function of time due to mass loss by stellar winds. 
{\it Right panel}:The black points show the Ne and C abundances observed at the surface
of WC stars by Dessart et al. (2000; filled circles) and by Crowther et al. (2006; filled triangle).
The dotted line show the prediction for a 60 M$_\odot$ stellar model with $Z$=0.020 (Meynet \& Maeder 2003) and the dashed line for a 60 M$_\odot$ stellar model with $Z$=0.014. }
\label{xi60}
\end{figure}

Libraries of nuclear reaction rates are available in Angulo et al. (1999), Rauscher et al. (2002)
\footnote{they can be obtained at the following web sites:
http://pntpm.ulb.ac.be/Nacre/nacre.htm 
http://phys.llnl.gov/Research/RRSN/}.
Arnould \& Katsuma (2007) discuss some reactions of great astrophysical interest.
Let us recall that most of the rates of the nuclear reactions
intervening in stars rely either completely or partially on theoretical calculations
(see e.g. Rolfs \& Rodney 1988).
Effects of changes of some nuclear reactions in stellar models have been recently studied for instance 
by Weiss et al. (2005) in low and intermediate mass stars, Herwig et al. (2006) in AGB stars, Decressin et al. (2007) in massive stars. 

To our knowledge, the only example where a nuclear property has been deduced from an
astrophysical observation is when Hoyle (1954) deduced the existence of an excited state of the carbon nuclei from the cosmic abundance ratios of carbon, oxygen and neon (see the interesting paper by Maeder 1999 on the discovery of the $3\alpha$-reaction and also the nice history
of nuclear astrophysics written by Celnikier 2006). In most cases,  
it is difficult (if not impossible) to constrain the rates of nuclear reactions from observations of stars.
Indeed, nuclear reactions occur, hidden, in the central regions of stars. In order to be able to observe the effects of nuclear reactions, it is necessary that some process removes the outer layers
allowing thus the central regions (where the nuclear reactions have occurred) to be uncovered. 
This is what happens for the most massive stars in our solar neighborhood. These stars lose very large
amounts of material as a result of strong stellar winds triggered by radiation pressure. At a given stage,
due to the removal of the outer envelope, 
layers having belonged to the convective H-burning core are exposed to the surface and gives the observers the
opportunity to see the products of H-burning and sometimes, when the winds are strong enough,  even of He-burning. Let us precise that due to the very thick wind, it is not possible to observe the surface of these stars, but the measurement of the abundances in the winds gives access to the abundances that were at the surface.

Massive stars whose envelope has been stripped off are called Wolf-Rayet stars.
Wolf--Rayet stars are nearly evaporating stars and thus are wonderful objects
to illustrate the effects of mass loss on massive star evolution.
A recent review devoted to WR stars may be found in Crowther (2007). 


As recalled above, the Wolf-Rayet stars present at their surface
the results either of CNO-process\-sing (WN stars) or of the $3\alpha$-process (WC or WO stars).
The left panel of Fig.~\ref{xi60} shows the evolution of the surface abundances in a rotating and 
a non--rotating model of a 60 M$_{\odot}$ star (the initial rotational velocity at the equator
is indicated). Although the two models are different with respect
to their initial physical conditions (one is rotating fast the other is not rotating), they display
very similar abundances during the WNE phase. During that phase, the star has lost its whole original
H-rich envelope and we can observe at the surface of the star, deep layers whose composition has reached
CNO equilibrium values. The nuclear equilibrium CNO values are essentially model independent as already stressed a long time ago (Smith \& Maeder 1991).
Thus in that case, the resulting N/C and N/O ratios are mainly representative of the nuclear processes and allow us to check the chains of nuclear reactions occurring in such stars.  
The good agreement between the observed and predicted values of CNO equilibrium 
indicates the general correctness of our understanding of the CNO cycle and of the relevant
nuclear data. 

The abundances 
observed during the WC/WO phase well correspond to the apparition at the surface of He-burning products (see the review by Crowther 2007).
In particular the high overabundance of $^{22}$Ne at the surface of the
WC star predicted by He-burning reactions is well confirmed by the observations (Willis 1999; Dessart et al. 2000; Crowther 2006). Note that the abundance of $^{22}$Ne at this stage (WC) is an indication
of the initial CNO content of the star. Indeed, $^{22}$Ne comes mainly from the destruction of $^{14}$N at the beginning of the
helium burning phase, this $^{14}$N being the result of the transformation of carbon and oxygen into nitrogen operated by the CNO cycle during the core H-burning phase.
In that respect it is interesting to note that comparison between observed Ne/He ratio at the surface of WC star with models computed with Z=0.02 show that models over predict the Ne abundance, while models starting
with the solar abundances given by Asplund et al. (2005) give a much better fit as can be seen in the right panel of Fig.~\ref{xi60}. This tends to
confirm that massive star in the solar neighborhood have initial metallicities in agreement with the
Asplund et al. (2005) solar abundances.
Let us note that this overabundance of $^{22}$Ne at the surface of WC stars is not only an important confirmation of the nuclear reaction chains occurring during He-burning, but is also related to
the question of the origin of the material accelerated into galactic cosmic rays (see recent
measurements of the $^{22}$Ne/$^{20}$Ne ratio in cosmic rays in Binns et al. 2005) and to the weak
s process in massive stars since $^{22}$Ne is the source of neutrons in these stars.

Changes of the surface abundances can also occur as a result of internal mixing.
In that case it might be very delicate to deduce from observed surface abundances strong conclusions about the
nuclear processes, especially that the internal mixing process may distort the changes of abundances due to nuclear processes by transporting more efficiently some elements than others.
This is the reason why, stellar modelers needs accurate nuclear reaction rates which are a prerequisite to
make progresses not only in stellar physics, but also in stellar nucleosynthesis and in our understanding of the chemical evolution of the galaxies.


\end{document}